\documentclass[%
 reprint,
 amsmath,amssymb,
 aps,
pra,
]{revtex4-2}
\usepackage[dvipsnames]{xcolor}
\usepackage{graphicx}
\usepackage{enumitem}
\usepackage{siunitx}
\DeclareSIUnit\gauss{G}
\usepackage{booktabs}
\usepackage{multirow}
\usepackage{mathtools}
\usepackage{flushend}
\usepackage[switch]{lineno}
\usepackage[utf8]{inputenc}
\usepackage{soul,xcolor}

\makeatletter 
\renewcommand{\fnum@figure}{\textbf{Fig.~\thefigure}}
\makeatother
\usepackage{braket}

\usepackage{times}
\usepackage{bm}
\usepackage{amsmath}

\newcommand{\probP}{\text{I\kern-0.15em P}}

\begin{document} 

\title{High-Precision Phase Control of an Optical Lattice with up to 50 dB Noise Suppression}

\author{Kendall Mehling}
\email{kendall.mehling@colorado.edu}%
\affiliation{JILA \& Department of Physics, University of Colorado Boulder, Boulder, CO, 80309-0440}

\author{Murray Holland}
\email{murray.holland@colorado.edu}%
\affiliation{JILA \& Department of Physics, University of Colorado Boulder, Boulder, CO, 80309-0440}

\author{Catie LeDesma}
\email{catherine.ledesma@colorado.edu}
\affiliation{JILA \& Department of Physics, University of Colorado Boulder, Boulder, CO, 80309-0440}

\date{\today}%

\begin{abstract}

An optical lattice is a periodic light crystal constructed from the standing-wave interference patterns of laser beams. It can be used to store and manipulate quantum degenerate atoms and is an ideal platform for the quantum simulation of many-body physics. A principal feature is that optical lattices are flexible and possess a variety of multidimensional geometries with modifiable band-structure. An even richer landscape emerges when control functions can be applied to the lattice by modulating the position or amplitude with Floquet driving. However, the desire for realizing high-modulation bandwidths while preserving extreme lattice stability has been difficult to achieve. In this paper, we demonstrate an effective solution that consists of overlapping two counterpropagating lattice beams and controlling the phase and intensity of each with independent acousto-optic modulators. Our phase controller mixes sampled light from both lattice beams with a common optical reference. This dual heterodyne locking method allows exquisite determination of the lattice position, while also removing parasitic phase noise accrued as the beams travel along separate paths. We report up to 50~dB suppression in lattice phase noise in the 0.1~Hz -- 1~Hz band along with significant suppression spanning more than four decades of frequency. When integrated, the absolute phase diffusion of the lattice position is only 10~\r{A} over 10 s. This method permits precise, high-bandwidth modulation (above 50kHz) of the optical lattice intensity and phase. We demonstrate the efficacy of this approach by executing intricate time-varying phase profiles for atom interferometry. 

\end{abstract}

\maketitle 

\section*{Introduction}

Optical lattices have proven to offer pristine, versatile platforms for the storage and manipulation of a variety of species, including ultracold neutral atoms, molecules, and ions. As a consequence, optical lattices have been employed to investigate a broad range of fundamental physics topics such as quantum simulators~\cite{doi:10.1126/science.aal3837,doi:10.1126/science.1177838,Mazza_2012}, condensed matter systems~\cite{JESSEN199695, doi:10.1080/00018730701223200, denschlag2002bose}, cold atom inertial sensing~\cite{Gebbe2021, PhysRevLett.102.240403, PhysRevLett.114.100405, PhysRevResearch.6.043120, PhysRevResearch.7.013246, ledesma2024vector},  timekeeping~\cite{Takamoto2005,Bloom2014, kim2023evaluation},  quantum gas microscopes~\cite{Bakr2009, PhysRevLett.114.193001, Gross2021}, efficient atom cooling techniques~\cite{Perrin1998, PhysRevLett.84.439}, and  entanglement generation~\cite{sorensen1999spin, meiser2008spin}. Their widespread application in atomic physics has been driven by their ability to model a variety of fundamental many-body Hamiltonian systems, including the Hubbard model~\cite{Bakr2009}, systems with SU(N) symmetry~\cite{Cazalilla_2014}, and high-energy physics lattice gauge theories~\cite{Zohar2015-tm}. In these systems, fine control can be achieved over the model parameters, flexibility that is not typically available in the solid-state systems that they aim to simulate. Reconfiguring parameters is achievable through experimental access to quantities such as the lattice spacing, position, intensity, and even multidimensional structure.

An optical lattice is composed of an interference pattern and can be constructed by directing a monochromatic laser onto a mirror and reflecting the incident light back upon itself to generate a standing wave of light. This configuration possesses intrinsic phase stability since the retromirror determines the spatial origin for the lattice nodes and antinodes. Even so, thermal, mechanical, or light pressure fluctuations can result in the mirror's position varying and causing uncertainty in the lattice phase.  To mediate this, the mirror can be passively~\cite{kasevich1992measurement} or actively~\cite{10.1063/1.1149838} stabilized, or the motion of the mirror measured with precision accelerometers, which allows adverse effects to be removed in post-processing~\cite{Geiger2011-oh,PhysRevA.108.043305}.

Although the retromirror configuration lends itself well to creating stationary lattices, an inherent problem is that it can be difficult to move the position of the optical lattice at will. Allowing the lattice to move introduces a control function that can be incorporated into quantum design methods to allow a whole class of new systems to be simulated and studied. Moving the lattice can be achieved by mounting a piezoelectric device to the retromirror~\cite{PhysRevLett.100.043602, PhysRevA.86.033615}. Although this is a straightforward technique, piezoelectric actuators suffer from constrained range of motion (i.e., limited attainable phase shifts), comparably low modulation bandwidths ($\sim$10~kHz), and unwanted drifts with age and environmental exposure. 

A method to overcome these drawbacks involves removing the retromirror and instead overlapping two counterpropagating laser beams in a ``free-space'' configuration. While this design increases experimental complexity, fine control over the lattice position is obtained through independent tuning of the relative frequencies of the counterpropagating lasers. Either an acousto-optic modulator (AOM) or acousto-optic deflector (AOD) can be placed in each individual beam path prior to forming the lattice, allowing for precise, high-bandwidth modulation of the optical intensity and phase ($>$50~kHz). However, because the two lattice beams traverse independent paths before they are spatially overlapped to form the lattice, they may experience differential time-dependent phases~\cite{Ma:94}. The effect of table vibrations and thermal fluctuations, air turbulence, and acoustic noise can cause additional relative phase instabilities that degrade performance and are difficult to combat. 

\begin{figure*}[t!]
            \centering
    \includegraphics[width=1\textwidth]{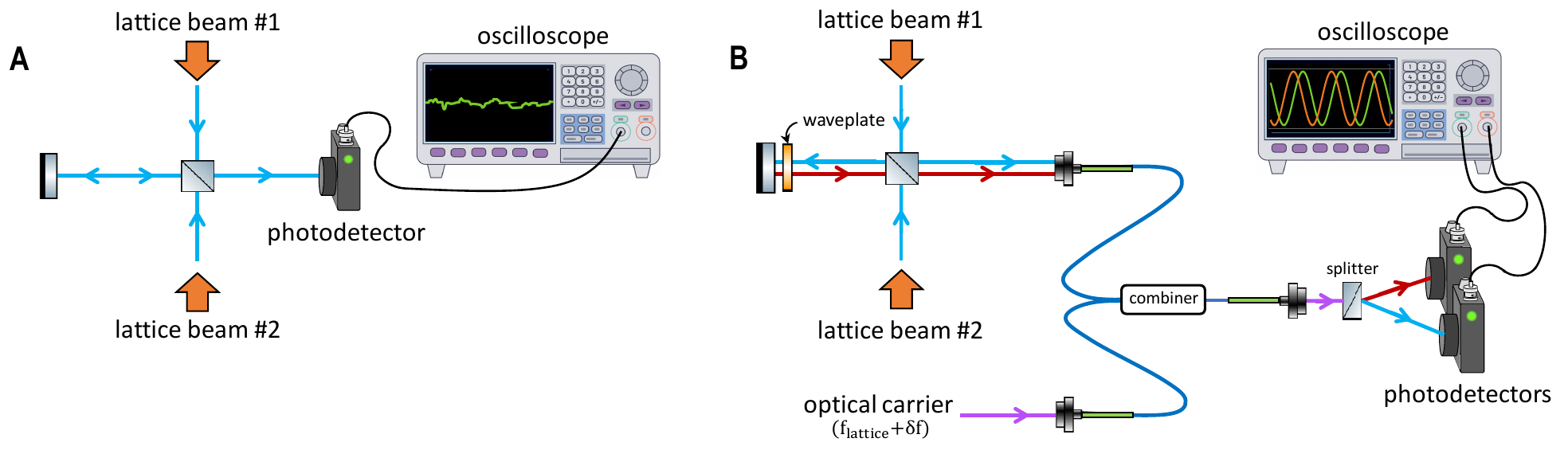}
    \caption{{\bf Lattice phase detection schemes.} (A) Direct phase detection scheme: Portions of lattice beam \#1 and \#2, seen in blue, are spatially overlapped onto a photodetector through the use of a beamsplitter and retromirror. (B) Dual heterodyne detection scheme: Portions of lattice beam \#1 and \#2 seen in blue are sampled and directed into a fiber coupler with orthogonal linear polarizations, denoted in blue and red. Each lattice beam of frequency $f_{\rm lattice}$ is shifted by an amount $\delta f$ with respect to a common optical carrier (purple) containing equal components of both polarizations. The lattice beam samples are combined with the optical carrier via a fiber combiner. At the fiber output, the beams are spatially discriminated and focused onto photodetectors for simultaneous heterodyne measurement. The measured photodiode beat notes monitor the phase of each lattice beam compared to the optical carrier and are used for feedback to stabilize the relative phase of the lattice.}
    \label{Heterodyne Concept}
\end{figure*}

In this paper, we present a solution that overcomes the challenges of achieving high-modulation bandwidths while maintaining extreme lattice phase stability for low-temperature many-body physics and precision metrology applications. We augment the free-space methodology with an advanced servo system that satisfies both the desired modulation capabilities and low-noise requirements. Mixing sampled lattice light with a single frequency-detuned optical carrier allows the measurement and control of the phase of each respective lattice beam~\cite{PhysRevApplied.9.034016}. By then frequency locking each beam to independent channels of an arbitrary waveform generator, we can apply intricate time-varying phase profiles while removing any parasitic phase noise due to the lattice beams transiting alternate paths.

This paper is organized as follows. We begin with a description of traditional phase detection techniques and introduce a dual heterodyne phase control concept. We then describe the optical schematic and accompanying electronic feedback systems. The phase controller's performance is then characterized through studies of the lattice phase diffusion and applied high-frequency modulation protocols. We finish with a discussion of future applications and extensions to this work.

\section*{Phase Controller Concept}

\begin{figure*}[t!]
    \centering
    \includegraphics[width=1\textwidth]{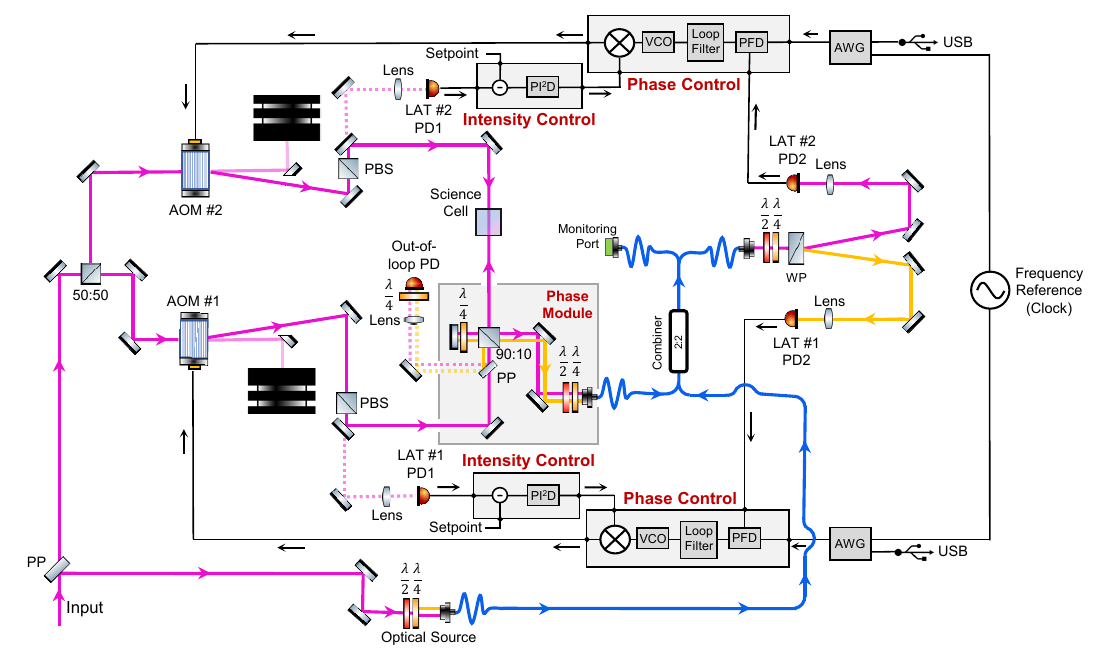}
    \caption{ {\bf Opto-electrical design of the optical lattice phase controller.} A single high-powered 1064~nm laser beam (input, lower left) is initially sampled and coupled into an optical fiber to serve as a common-mode optical source. The remaining input light is equally split to create two optical lattice beams that pass through their own respective AOMs. Each AOM is driven by an RF signal that control the intensity and phase of each lattice beam. Within the schematic, optics and electronics are abbreviated as follows: pickoff plate (PP), photodiode~(PD), lattice~(LAT), polarizing beamsplitter (PBS), Wollaston prism (WP), voltage controlled oscillator (VCO), proportion-integral-differential (PID), phase frequency detector (PFD), arbitrary waveform generator (AWG), and universal serial bus (USB). The out-of-loop photodiode is not an essential part of the controller but allows direct optical interferometric measurement of the lattice interference pattern. The science cell is a glass chamber at ultrahigh vacuum for ultracold atom experiments.}
    \label{Schematic}
\end{figure*}

An elementary method for measuring the relative phase of a free-space optical lattice can be accomplished by sampling light from each constituent beam and overlapping them onto a single photodetector. This method, depicted in Fig.~\ref{Heterodyne Concept}A, uses a beamsplitter and mirror combination to redirect lattice beam~\#2 (LAT \#2) to overlap with lattice beam~\#1 (LAT \#1) on a detector to reveal variations in the phase of the interference pattern. With this measurement, a feedback scheme could be constructed to stabilize the lattice phase to a fixed value, resulting in a low-frequency, direct comparison error signal. However, due to its susceptibility to long-term drift and low-frequency noise, this method is not ideal for high precision applications. Furthermore, there are no straightforward solutions that also allow rapid, piezo-free phase modulation to be incorporated into the feedback scheme. 

A better approach is to draw inspiration from traditional techniques used to control laser frequencies for ultrastable metrology experiments. Well-established methods such as the Pound–Drever–Hall (PDH) technique~\cite{Drever1983-jc}, offset laser locking schemes~\cite{Corwin:98, Thorpe:08}, and the use of frequency combs~\cite{RevModPhys.78.1279,RevModPhys.78.1297, RevModPhys.75.325, Diddams:10, Picqué2019} for frequency stabilization all involve combining laser light with an optical reference to generate a beatnote that can function as an error signal for noise correction. Because this signal tends to be on the order of MHz to GHz, an abundance of radio frequency (RF) electronics are well established for creating custom servo solutions for precise frequency manipulation. We leverage these previous concepts in our design by combining the lattice beams with an optical reference, or carrier, to generate RF beatnotes for independent frequency locking. These beatnotes allow fine control of the frequency of each lattice beam and therefore the relative phase of the optical lattice pattern.

As illustrated in the dual heterodyne scheme seen in  Fig.~\ref{Heterodyne Concept}B, a beamsplitter is used to sample light from each lattice beam. However, a quarter-waveplate inserted in front of the mirror rotates the polarization of LAT~\#2 so that the sampled lattice light now consists of two distinct linear polarizations, each uniquely associated with a single lattice beam. When coupled into a polarization maintaining (PM) fiber combiner, these two polarized beams are then mixed with an optical carrier that contains an equal superposition of both linear polarization states and is frequency shifted from the lattice frequency, $f_{\rm lattice}$, by an amount $\delta f$. At the output of the PM fiber combiner, the two polarization modes are spatially separated and focused onto independent photodetectors. The intensity measured by the independent photodetectors encodes the time-varying phase of each lattice beam onto a high-frequency carrier oscillating at $\delta f$. Comparison of each beatnote to a stable RF reference provides a suitable error signal for independent feedback to modify the frequency of each lattice beam.

An essential advantage of this approach is that any phase noise added to the optical carrier is common mode and therefore doesn't contribute to the relative phase of the lattice. Without this property, acoustic and other fiber noise would be prohibitive to the operation of the device. Furthermore, with this scheme it is straightforward to introduce a modulation to the position of the optical lattice. This is achieved by updating the relative RF drive that each photodiode beatnote is stabilized to. By using time-dependent arbitrary RF waveforms to drive either lattice AOM, the position of the nodes and antinodes of the optical lattice can be dynamically controlled. Note that, although the lattice beams are independently stabilized to separate RF references, they can be locked together by simply synchronizing both references with a single clock, yielding a stable relative phase.

\section*{Experimental Setup}

The schematic of our counterpropagating lattice design and accompanying intensity and phase stabilization components is illustrated in Fig.~\ref{Schematic}. The glass science cell depicts the location of the optical lattice for ultracold atom experiments. 

Initially, a small portion ($\sim$4\%) of laser light generated by a IPG 1064nm laser (YLR-1064-30W) is split off by a pick-off plate (PP) to serve as the optical carrier. The amount of power remaining in the main beam varies from 1-5~W depending on what beam waists and achievable lattice depths are desired in the experiment. The remaining optical power is then split by a 50:50 beam splitter (BS) to generate two distinct lattice beams, each of which are independently controlled by separate AOMs, which have 80~MHz center frequencies. After diffracting through an AOM, the polarization of each beam is then filtered by a polarizing beam splitter before reflecting off of two additional mirrors that direct the beams into the cell to form the lattice interference pattern.

To counteract parasitic phase noise that may be separately imprinted on the lattice beams during their transit to the cell, the dual heterodyne phase controller scheme described in Fig.~\ref{Heterodyne Concept}B is employed. The optics needed to sample the lattice light can be seen in the phase module depicted in Fig~\ref{Schematic}. Since we cannot place optics inside the science cell, light from each lattice beam is instead sampled as close as feasible to the cell using a 90:10 beamsplitter. Because the lattice beams traverse the phase module in opposite directions, they get reflected $180^{\circ}$ from one another at the beamsplitter.  A mirror and quarter-waveplate serve to both redirect and rotate the polarization of LAT \#1 so that after LAT \#1 passes a second time through the beamsplitter, the light extracted from both lattice beams now travel collinearly with orthogonal linear polarizations. 

Alignment of the mirror pair and retroreflecting mirror following the 90:10 beamsplitter allows up to 5\% of each lattice beam to be coupled into an input port of a 2:2 polarization maintaining fiber (PNH1064R5A2). The $\lambda/4$, $\lambda/2$ waveplate pair before the fiber coupler are adjusted to align the polarization of LAT \#2 to one of the polarization axes of the fiber. Adjustment of the retroreflecting mirror and rotation of its attached $\lambda/4$ waveplate align the polarization of LAT~\#1 to the orthogonal polarization axis of the fiber. We emphasize that for proper operation of the phase controller, it is critical that the two lattice beams are cleanly coupled into orthogonal polarization modes of the PM fiber. The polarization extinction ratios (PERs) along each axis of the PM fiber are measured via a bench-top PER meter (ERM-200) and can be set to be greater than 33 dB, which surpasses the typically quoted fiber extinction ratio of 25--30 dB. 

The optical carrier light that was sampled prior to splitting the light into separate lattice paths is fiber coupled into the other input port of the 2:2 PM fiber coupler to be combined with the lattice light. Transport within the 2:2 fiber coupler enables straightforward delivery away from the space-constrained experimental platform and ensures mixing of each sampled lattice beam with the common optical carrier. Of the two fiber coupler output ports, one port is used to monitor the transmitted optical power and polarization mode of each of the three beams. Power emitted from the other output port is used in the feedback scheme for phase control. The light from the fiber is collimated and directed through a pair of waveplates followed by a Wollaston prism (WP10-B) that spatially splits the two linear polarization states with a 20$^\circ$ separation angle. The waveplates at the fiber output are oriented to ensure complete discrimination between lattice beams following the prism, which has a quoted polarization extinction ratio exceeding 100,000:1. The common optical carrier contains an equal superposition of both linear polarization modes of the fiber ensuring that it is equally split along both paths following the Wollaston prism. The spatially separated beams are then steered and focused onto high speed photodiodes (PDA05CF2) denoted PD~\#1 and PD~\#2. These photodiodes measure the beat frequency ($\delta f \approx 80$~MHz) and relative phase between the lattice beams and the optical carrier. 

Each independent heterodyne signal is subsequently passed to a phase lock loop (PLL) electronic system. The lattice photodiode outputs are AC coupled through a bias tee (ZX85-12-S+) and then passed through a limiting amplifier (AD8306) which acts to clamp the AC signal. The limiting amplifier mitigates the effects of amplitude fluctuations and provides suitable gain for phase locking, even with $<$100~$\mu$W of light transmitted through the 2:2 fiber coupler.

The clamped beatnote signal is then passed to a low-phase noise phase-frequency detector (PFD, HMC439) that compares the phase/frequency of the processed photodiode signal to the RF reference. The PFD differential output is then sent to a high speed $\text{PI}^2\text{D}$ loop filter which rectifies the input PFD pulse trains into a control voltage that is used to adjust the VCO (ZX95-78+) RF frequency that drives each lattice AOM. To implement complex phase profiles, two channels of an arbitrary waveform generator (Keysight 33622A) are used as RF references for each lattice PLL. A 10 MHz rubidium frequency standard (FS725) is used as a common reference clock for the RF generator and to synchronize all measurement equipment. 

Although not directly related to phase control, our system contains additional capabilities. When the phase controller is not engaged, each lattice AOM frequency is instead controlled directly by the arbitrary waveform generator used as the RF reference. A series of splitters and switches (not shown in Fig.~\ref{Schematic}) are used to alternate between the two modes of operation. The intensity of each beam is also independently stabilized via a feedback system using light that is picked off from the lattice beams prior to the phase module. As seen in Fig \ref{Schematic}, following each AOM, leakage light transmitted through a mirror along each path is steered and focused onto a photodiode (PDA10A). To actively stabilize the beam intensity, the photodiode voltage is compared to a computer-controlled setpoint using home-built AOM intensity servos. The proportional-integral (PI$^{2}$) loop filter subsequently outputs a control voltage to the IF port of an RF mixer (ZAD-3) which actuates the RF power passed to each lattice AOM. 

\begin{figure*}[t!]
    \centering
    \includegraphics[width=.7\textwidth]{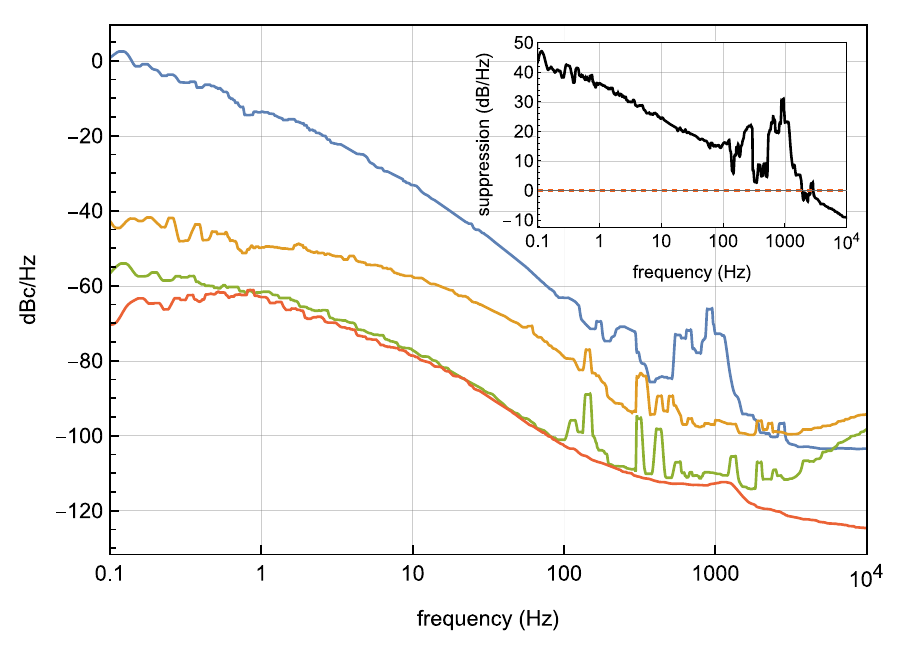}
    \caption{ {\bf Phase noise characteristics of the optical lattice phase controller}. The measured out-of-loop phase noise for the free-running~(blue) and phase-locked lattice (orange), with the suppression ratio displayed in the inset. Phase noise suppression is achieved over 4 decades of frequency, with particular enhancement seen in the frequency band of 0.1~Hz -- 100~Hz. For comparison, the phase noise of the 80 MHz RF reference (red) and an in-loop photodetector heterodyne signal (green) are measured over the same frequency range. }
    \label{phase noise}
\end{figure*}

\vspace*{1pc}
\section*{Lattice Controller Characterization}

Evaluation of the performance of the phase controller is achieved by measuring the relative phase of the optical lattice using an out-of-loop (OOL) photodiode (PDA05CF2). This can be seen as part of the phase module in Fig.~\ref{Schematic}, where after the sampled  LAT \#1 light  is retro-reflected by the mirror, 10\% of its power again reflects off the 90:10 beamsplitter so that a fraction of LAT \#1 light now copropagates along the same path as LAT \#2. Both beams are then sampled by a pickoff plate and then focused onto the OOL photodiode where a quarter waveplate placed before the diode serves to combine the orthogonal polarizations states of the two lattice beams for interference.

\subsubsection*{Phase Noise Suppression}

To characterize the noise suppression performance of the phase controller, a frequency difference is imparted between the lattice beams. The stability of the resulting beatnote signal detected on the OOL photodiode is then measured using a cross-correlation phase noise analyzer (R\&S FSWP26). Generally a 1~MHz offset is used for characterization of the controller but we note that up to a 5~MHz frequency shift can be imparted between the two lattice AOM drive frequencies while still maintaining precise control of the lattice phase.

 Several phase noise traces of relevant signals in the phase feedback system can be seen in Fig~\ref{phase noise}. For each signal, the phase noise analyzer trace indicates the power relative to the carrier~(dBc) for a range of offset frequencies. The free-running lattice phase noise, detected on the OOL photodiode signal for a 1~MHz offset frequency is shown in blue and displays substantial phase noise, particularly in the 0.1~Hz -- 1~kHz range. This can be attributed to known effects such as acoustic vibrations of the experimental enclosure, pneumatic control of the optics table, and/or ambient vibrations of the concrete substrate on which the experiment stands. The phase noise of the 1~MHz signal with the phase controller engaged is depicted in orange and  relative to the free-running trace, substantial noise suppression can be seen over the same frequency band. 

The inset of Fig~\ref{phase noise} compares the two cases by illustrating the noise ratio between the phase-locked and free-running lattice. Near 0.1~Hz we achieve almost 50~dB of observed noise reduction, with sizable noise suppression also seen over more than four orders of magnitude of frequency. The integrated phase noise over the entire depicted frequency range for the free-running and phase-locked beatnotes are $-5.40$~dBc (757~mrad) and $-40.6$~dBc (13.1~mrad), respectively. Although we cannot measure the lattice phase in the vacuum of the science cell, we emphasize that the actual noise reduction is expected to be slightly better at this location. This is due to added noise that arises from propagation through the cell, within the phase module, and to the OOL photodiode where the measurement is taken.

\begin{figure*}[t!]
    \centering
    \includegraphics[width=.95\textwidth]{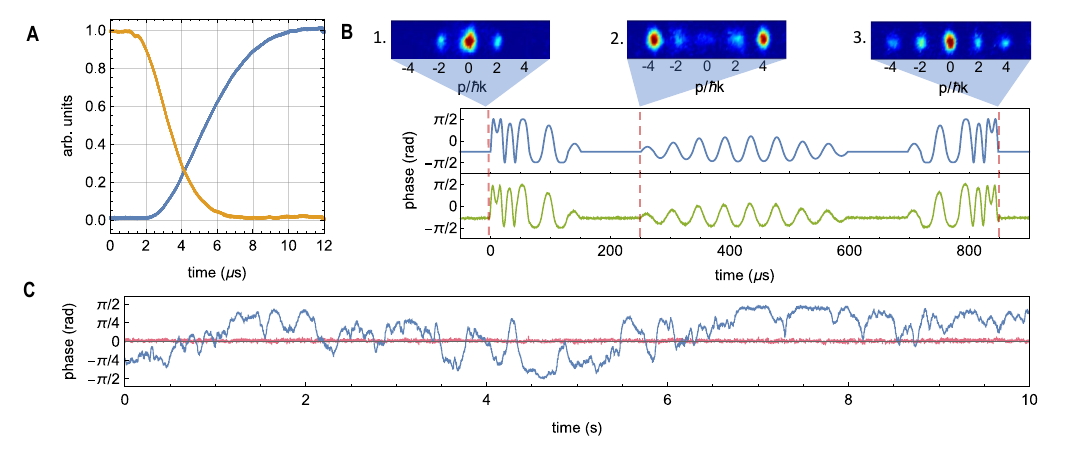}
    \caption{ {\bf  Performance characteristics of the optical lattice phase controller.} (A) Phase (orange) and intensity (blue) response curves of the feedback systems for small impulse changes to their lockpoints. (B) High-fidelity execution of a Bloch-band interferometry (BBI) sequence~\cite{PhysRevResearch.6.043120, PhysRevResearch.7.013246, ledesma2024vector} confirmed via both absorption images, following (1) the initial lattice load, (2) beamsplitter, and (3) recombination. The measured lattice phase (green, bottom) is compared to the ideal phase control sequence (blue, top). (C) Free-running (blue) and phase-locked (red) time series of the lattice phase over 10~s with a static phase applied between the RF references. }
    \label{dynamic control}
\end{figure*}

Along with the OOL photodiode phase noise traces, the phase noise of the RF reference (red) and a stabilized in-loop optical heterodyne signal (green) are also depicted. For clarity only one of the two heterodyne traces is shown since both are observed to be comparable. Each trace characterizes the phase noise of an 80~MHz signal, which is the center frequency for both lattice AOMs. While actively stabilized, the phase noise of the in-loop photodiode heterodyne signal closely matches the RF reference from 0.1~Hz -- 2~kHz. We attribute the increased phase noise of the in-loop heterodyne signal at these frequencies to minute deviations in the lattice beam polarization which can lead to minor crosstalk between the heterodyne signals and reduced feedback performance. 
At frequencies greater than 1~kHz, the increased phase noise of the in-loop signal is a result of additive phase noise of the PLL for offset frequencies which approach the feedback bandwidth.

\subsubsection*{High-Bandwidth Modulation}

In addition to greatly reducing relative phase noise of the optical lattice, this phase control design also permits precise, high-bandwidth modulation of the intensity and phase of the optical lattice. In Fig.~\ref{dynamic control}A, we show the response curves for both intensity (blue) and phase (orange) for small impulse changes of the desired lock point.  The modulation bandwidths for the intensity and phase of the optical lattice are determined to be approximately 70~kHz, and 200~kHz respectively. 

The flexibility of this phase controller for high bandwidth modulation opens the door to a rich landscape of quantum simulation applications. As an example of this, we demonstrate an implementation of Bloch-band atom interferometry~(BBI), that is, atom interferometry performed entirely within the valence band and conduction band of an optical lattice system. It has recently been shown that by modulating the phase of an optical lattice, Bose-Einstein condensed atoms can be manipulated to transit between the Bloch eigenstates and thereby be made to undergo the traditional steps of two-path interferometry. A matterwave accelerometer~\cite{PhysRevResearch.6.043120}, measurements of both the magnitude and direction of a vector acceleration~\cite{ledesma2024vector}, and presentation and demonstration of a universal gateset for atom interferometry~\cite{PhysRevResearch.7.013246} have all been performed using this method. In BBI, the phase profiles needed to realize each atom optic component are determined via machine-learning~\cite{PhysRevResearch.3.033279} which results in non-intuitive time-varying solutions that can be quite complex, requiring high-bandwidth control of the lattice position. Furthermore, to advance this precision measurement technology, phase stability of the optical lattice is also essential, making it an obvious candidate for this type of phase controller.  

As seen in Fig \ref{dynamic control}B, a representative  BBI sequence is applied to atoms loaded into an optical lattice that is controlled via the presented phase controller scheme. For this example, a cloud of approximately 30,000 Bose-Einstein condensed rubidium atoms were used to validate the application of the interferometry protocol, which consists of a beamsplitter, mirror, and recombination step. The theoretical phase profile of the full sequence can be seen in blue within Fig.~\ref{dynamic control}B and was digitized at a sampling rate of 50~ns and then superimposed onto an 80~MHz carrier to create the referenced arbitrary waveform for the phase controller.  The phase of the optical lattice as measured on the OOL photodetector with the phase controller engaged can be seen in green, demonstrating superb agreement with the ideal phase profile. We note that the entire interferometry sequence is less than 1~ms in duration, and the measured phase closely matches the intricate features of the machine-learned atom-optic components. 

In this example, the atoms themselves can also be used to validate the performance of the applied interferometry sequence and thus the phase controller. The absorption image pop-outs (1-3) of Fig.~\ref{dynamic control}B, taken in separate experiments after the atoms have undergone 12~ms of time of flight, depict the momentum decomposition of the atomic ensemble at three different points in the protocol. In (1) the atoms are seen to be initially occupying the ground state of the lattice and therefore are embedded deeply in the valence band. In (2) the atoms have been transformed by an atom-optic beamsplitter into a conduction band state that manifests as a superposition of atoms traveling in two opposite directions. In (3), after reflection and application of the atom-optic recombiner, the atoms are observed to be once again in the original valence band ground state. Here the experimental fidelities of each transformation exceed 90\%. 

\subsubsection*{Low Frequency Phase Noise Suppression}

Engagement of the phase controller also provides significant suppression of low frequency phase noise in the stabilized lattice. To illustrate the long-term stability obtained by the feedback system, a fixed phase (i.e no frequency detuning) was set between the RF references driving the lattice AOMs and the relative phase of the lattice was then monitored on the OOL photodiode. The free-running (blue) and the locked phase (red) of the optical lattice for a 10~s time series can be seen in Fig.~\ref{dynamic control}C. During the dynamical evolution, the $\sim 50$~dB suppression previously measured as a function of frequency in Fig.~\ref{phase noise} is clearly observed, but here demonstrated in the time domain. Large variations of the lattice phase that are evident when the phase controller is not engaged are seen to be effectively removed when the phase controller is active. As previously stated, with the lattice phase locked, the integrated phase noise is 13.1 mrad in the frequency range 0.1~Hz~--~10~kHz. This implies that over the 10 s measurement interval, the absolute lattice rms positional uncertainty due to lattice phase diffusion is of order 10~\r{A}. 

\section*{Conclusion and Discussion}

In this paper, we have presented a free-space optical lattice design and accompanying intensity and phase servos which considerably reduce phase noise while also enabling precise, high-bandwidth modulation of the lattice intensity and phase. Performance of the phase controller was validated through an out-of-loop measurement of the lattice phase noise, which shows up to 50~dB suppression between 0.1~Hz -- 1~Hz, along with significant suppression over more than four orders of magnitude of frequency. The control system reduces the relative phase noise of completely independent optical paths down to a single retroreflecting mirror/beamsplitter combination that acts as the lattice's inertial reference. The high-fidelity execution of rapid BBI gates to ultracold atoms in an optical lattice was confirmed via absorption images of the atomic momentum decomposition and monitoring of an out-of-loop photodiode. This example highlights the dynamic programmability of the phase controller for quantum design solutions using ultracold atoms in optical lattices.

At the present, the discrepancy in phase noise suppression between the OOL photodiode signal and the in-loop heterodyne signals can be attributed to unwanted vibrations of the retroreflecting mirror/beamsplitter or to temporal fluctuations of the air paths in the phase module and between the phase module and the science cell. However, as previously mentioned, significant attention has been focused on the stabilization of optical lattice retromirrors via passive vibration isolation, active position feedback, or direct measurement of the displacement of the inertial reference for post processing.  Employing these techniques alongside the presented optical lattice phase controller could enable further suppression of unwanted lattice phase noise. An increase in performance could also be achieved by locking the lattice beams to a more stable RF reference than is currently provided by the arbitrary waveform generators. 

Lastly, although the lattice scheme seen in Fig.~\ref{Schematic} is of free-space design, it should also be noted that the phase controller possesses the ability to suppress phase noise incurred via transport of light through optical fibers. Such acoustic noise lies well within the modulation bandwidth of the phase control system \cite{Ma:94}, allowing the feedback to cancel any deleterious noise that is added through fiber transport. As quantum gas experiments progress toward field-deployable architectures with limited footprints, it is critical to adopt experimental designs that reduce size, weight and power (SWaP). This strongly motivates light distribution solutions using optical fibers. Transport of laser light within distinct optical fibers also offers enhanced spatial mode quality, reduced physical size, and can simplify optical alignment.

\section*{Acknowledgments}
We thank T. Brown for many useful discussions and assistance in characterizing both the JILA home-built loop filters and AOM intensity controllers. This research was supported in part by NASA under grant number 80NSSC23K1343, NSF PFC 2317149; and NSF OMA 2016244. MH acknowledges support from the University of Coloroado Boulder OEDIT partnership.

\bibliography{ref}

\end{document}